\documentclass[useAMS,usenatbib]{mn2e}

\usepackage{graphicx}

\def\aV{\mbox{$\rm A_V$}}

\def\jh{\mbox{$(J-H)$}}
\def\hk{\mbox{$(H-K_s)$}}
\def\jk{\mbox{$(J-K_s)$}}
\def\mMJ{\mbox{$(m-M)_J$}}
\def\mMo{\mbox{$(m-M)_O$}}
\def\ebv{\mbox{$E(B-V)$}}

\def\ejh{\mbox{$E(J-H)$}}

\def\rc{\mbox{$R_{\rm c}$}}

\def\rl{\mbox{$R_{\rm RDP}$}}
\def\rx{\mbox{$R_{\rm ext}$}}

\def\ms{\mbox{$M_\odot$}}
\def\ds{\mbox{$d_\odot$}}
\def\rs{\mbox{$R_\odot$}}
\def\dgc{\mbox{$R_{\rm GC}$}}

\def\jj{\mbox{$J$}}
\def\hh{\mbox{$H$}}
\def\ks{\mbox{$K_s$}}

\def\ns{\mbox{$N_{1\sigma}$}}
\def\no{\mbox{$N_{\rm obs}$}}
\def\nc{\mbox{$N_{\rm cl}$}}
\def\sFS{\mbox{$\rm\sigma_{FS}$}}
\def\fsU{\mbox{$FS_{\rm unif}$}}

\title[The very young OCs NGC\,2244 and NGC\,2239]{Probing the age and structure of the
nearby very young open clusters NGC\,2244 and NGC\,2239}

\author[C. Bonatto and E. Bica]{C. Bonatto$^1$\thanks{E-mail: charles@if.ufrgs.br} and 
E. Bica$^1$\thanks{E-mail: bica@if.ufrgs.br}\\
$^1$ Departamento de Astronomia, Universidade Federal do Rio Grande do Sul\\ 
Av. Bento Gon\c{c}alves 9500, Porto Alegre 91501-970, RS, Brazil}

\begin{document}

\pagerange{\pageref{firstpage}--\pageref{lastpage}}

\maketitle

\label{firstpage}

\begin{abstract}
The very young open cluster (OC) NGC\,2244 in the Rosette Nebula was studied with 
field-star-decontaminated 2MASS photometry, which shows the main-sequence (MS) 
stars and an abundant pre-MS (PMS) population. Fundamental and structural parameters
were derived with colour-magnitude diagrams (CMDs), stellar radial density profiles 
(RDPs) and mass functions (MFs). Most previous studies centred NGC\,2244 close to 
the bright K0V star 12\,Monocerotis, which is not a cluster member. Instead, the 
near-IR RDP indicates a pronounced core near the O5 star HD\,46150. We derive an 
age within 1---6\,Myr, an absorption $\aV=1.7\pm0.2$, a distance from the Sun 
$\ds=1.6\pm0.2$\,kpc ($\approx1.5$\,kpc outside the Solar circle), an MF slope 
$\chi=0.91\pm0.13$ and a total (MS$+$PMS) stellar mass of $\sim625\,\ms$. Its RDP 
is characterised by the core and cluster radii $\rc\approx5.6\arcmin$ ($\approx2.6$\,pc) 
and $\rl\approx10\arcmin$ ($\approx4.7$\,pc), respectively. Departure from dynamical 
equilibrium is suggested by the abnormally large core radius and the marked central 
stellar excess. We also investigate the elusive neighbouring OC NGC\,2239, which is 
low-mass ($m_{MS+PMS}\approx301\,\ms$), young ($5\pm4$\,Myr) rather absorbed 
($\aV=3.4\pm0.2$), and located in the background of NGC\,2244 at $\ds=3.9\pm0.4$\,kpc. 
Its RDP follows a King-like function of $\rc\approx0.5\arcmin\approx0.5$\,pc and $\rl\approx5.0\arcmin\approx5.6$\,pc. The MF slope, $\chi=1.24\pm0.06$, is essentially 
Salpeter's IMF. NGC\,2244 is probably doomed to dissolution in a few $10^7$\,yr. 
Wide-field extractions and field-star decontamination increase the stellar statistics 
and enhance both CMDs and RDPs, which is essential for faint and bright star clusters.
\end{abstract}


\section{Introduction}
\label{Intro}

Still in the process of emerging from the parent molecular cloud, star clusters
younger than about 5\,Myr usually present a developing main sequence (MS) and a
significant population of pre-MS (PMS) stars. However, depending on the initial 
cluster mass, star-formation efficiency and mass of the more massive stars, the 
rapid early gas removal (from supernovae and massive star winds) may impart important 
changes to the original gravitational potential. One consequence of the reduced potential 
is that stars, especially the low mass ones, moving faster than the scaled-down escape 
velocity may be driven into the field. Over a time-scale of $10-40$\,Myr,
this effect can dissolve most of the very young star clusters (e.g. \citealt{GoBa06}). 
Indeed, estimates (e.g. \citealt{LL2003}) predict that only about 5\% of the 
embedded clusters are able to dynamically evolve into bound open clusters (OCs). 

On observational grounds, the dramatic changes in the potential affecting the early 
cluster spatial structure should be reflected on the stellar radial density profile 
(RDP). Bochum\,1 (\citealt{BBD2008}), for instance, can be an example of this scenario, 
in which the irregular RDP does not follow a cluster-like profile. This suggests significant 
profile erosion or dispersion of stars from a primordial cluster. 

In the present paper we address the case of NGC\,2244 in the Rosette Nebula, which is also 
related to the Monocerotis\,OB2 association (e.g. \citealt{Zuniga08}). Historically, in 
colour-magnitude diagram (CMD) studies some authors centred the large-scale structure on 12\,Mon, 
which is a bright foreground star of spectral type K0V. When only wide-field CMDs are considered, 
the adoption of this centre does not affect the results. However, as will be explored in this work 
in the context of investigating the cluster structure, that region is definitely at the cluster 
periphery.

Based on Shanghai Observatory plates with baselines of 34 and 87\,yr, \citet{CdGZ07}
derived proper motions (PMs) and  membership probabilities for NGC\,2244. They found mass 
segregation, but no velocity-mass dependence, indicating a primordial mass segregation related 
to the star-formation process. With arguments based on published initial mass functions (IMFs)
and the measured internal velocity dispersion of $\rm\approx35\,km\,s^{-1}$, they concluded that
NGC\,2244 will be dissolved on a short time-scale.

Additionally, in the area of the Rosette Nebula, the cluster candidate NGC\,2239 has been 
frequently included in catalogues, but hardly studied. Both NGC\,2244 and NGC\,2239 are optical 
clusters, while the area includes numerous infrared embedded clusters in the Rosette Molecular 
Cloud (e.g. \citealt{PheLad97}).

This work employs 2MASS\footnote{The Two Micron All Sky Survey, All Sky data release
(\citealt{2mass1997}), available at {\em http://www.ipac.caltech.edu/2mass/releases/allsky/}} 
near-IR \jj, \hh, and \ks\ photometry. The 2MASS spatial and photometric uniformity allow 
extractions of wide surrounding fields that provide high star-count statistics. This property 
makes 2MASS an excellent resource to extract photometry of a broad variety of star clusters, 
the wide field ones in particular. For this purpose we have been developing quantitative 
tools to statistically disentangle cluster evolutionary sequences from field stars in 
CMDs. Decontaminated CMDs have been used to investigate the nature 
of cluster candidates and derive their astrophysical parameters (e.g. \citealt{ProbFSR}). 
Basically, we apply {\em (i)} field-star decontamination to 
measure the statistical significance of the CMD morphology, which is fundamental to derive
reddening, age, and distance from the Sun, and {\em (ii)} colour-magnitude filters, which are 
essential for intrinsic stellar RDPs, as well as luminosity and mass 
functions (MFs). In particular, the use of field-star decontamination in the construction of 
CMDs has proved to constrain age and distance more than working with raw (observed) 
photometry, especially for low-latitude OCs (\citealt{discProp}).

2MASS can be deep for nearby young or old OCs. For instance, our group has studied the young 
OCs NGC\,6611 (\citealt{N6611}) and NGC\,4755 (\citealt{N4755}). Abundant pre-MS (PMS) stars 
were seen in the $\approx1$\,Myr old NGC\,6611, which is essentially embedded, and a few 
remaining ones in the $\approx14$\,Myr old NGC\,4755. As nearby older OCs we cite NGC\,2477 
(\citealt{DetAnalOCs}) and M\,67 (\citealt{M67}).

\begin{figure*}
\begin{minipage}[b]{0.50\linewidth}
\includegraphics[width=\textwidth]{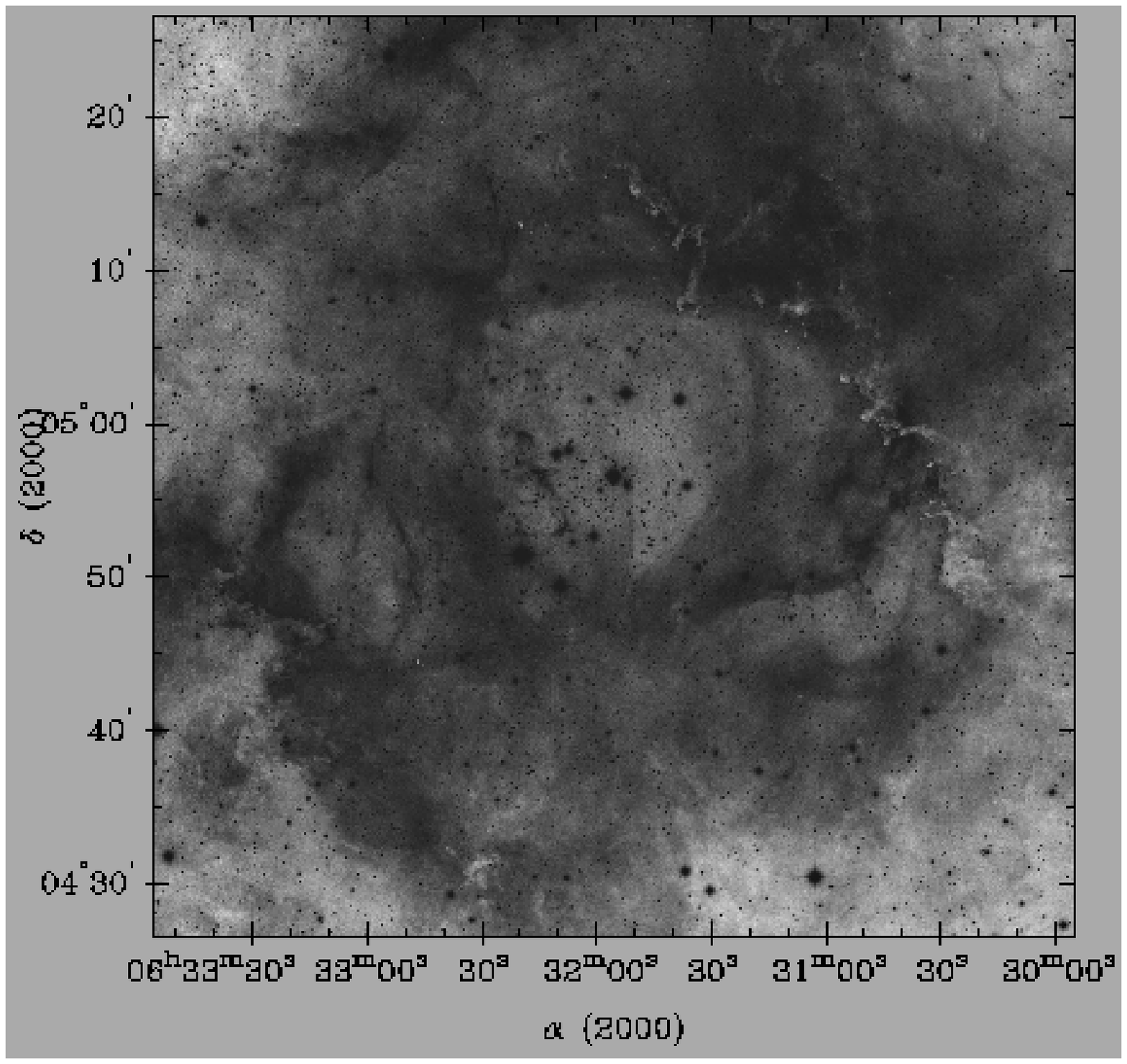}
\end{minipage}\hfill
\begin{minipage}[b]{0.50\linewidth}
\includegraphics[width=\textwidth]{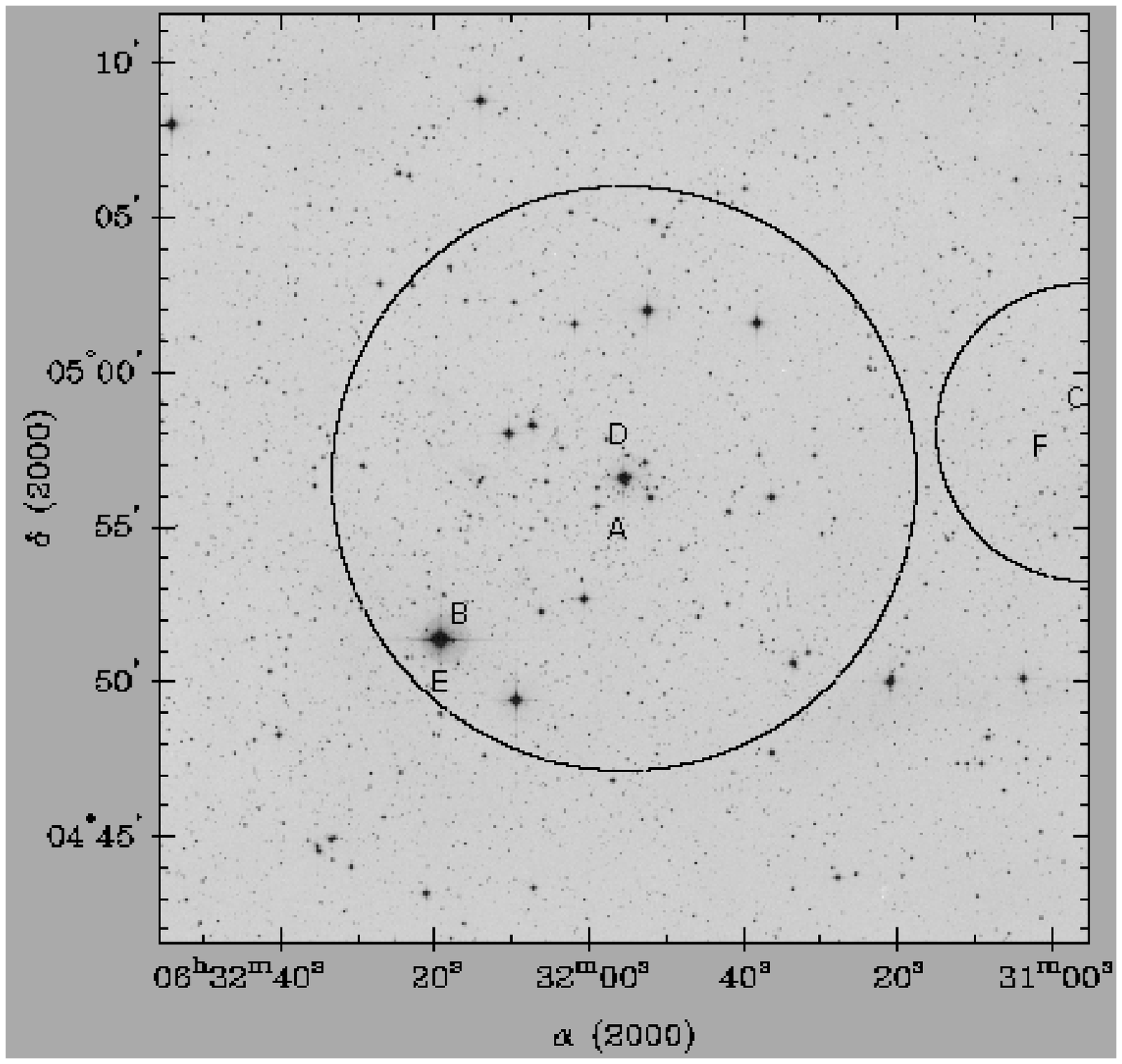}
\end{minipage}\hfill
\begin{minipage}[b]{0.50\linewidth}
\includegraphics[width=\textwidth]{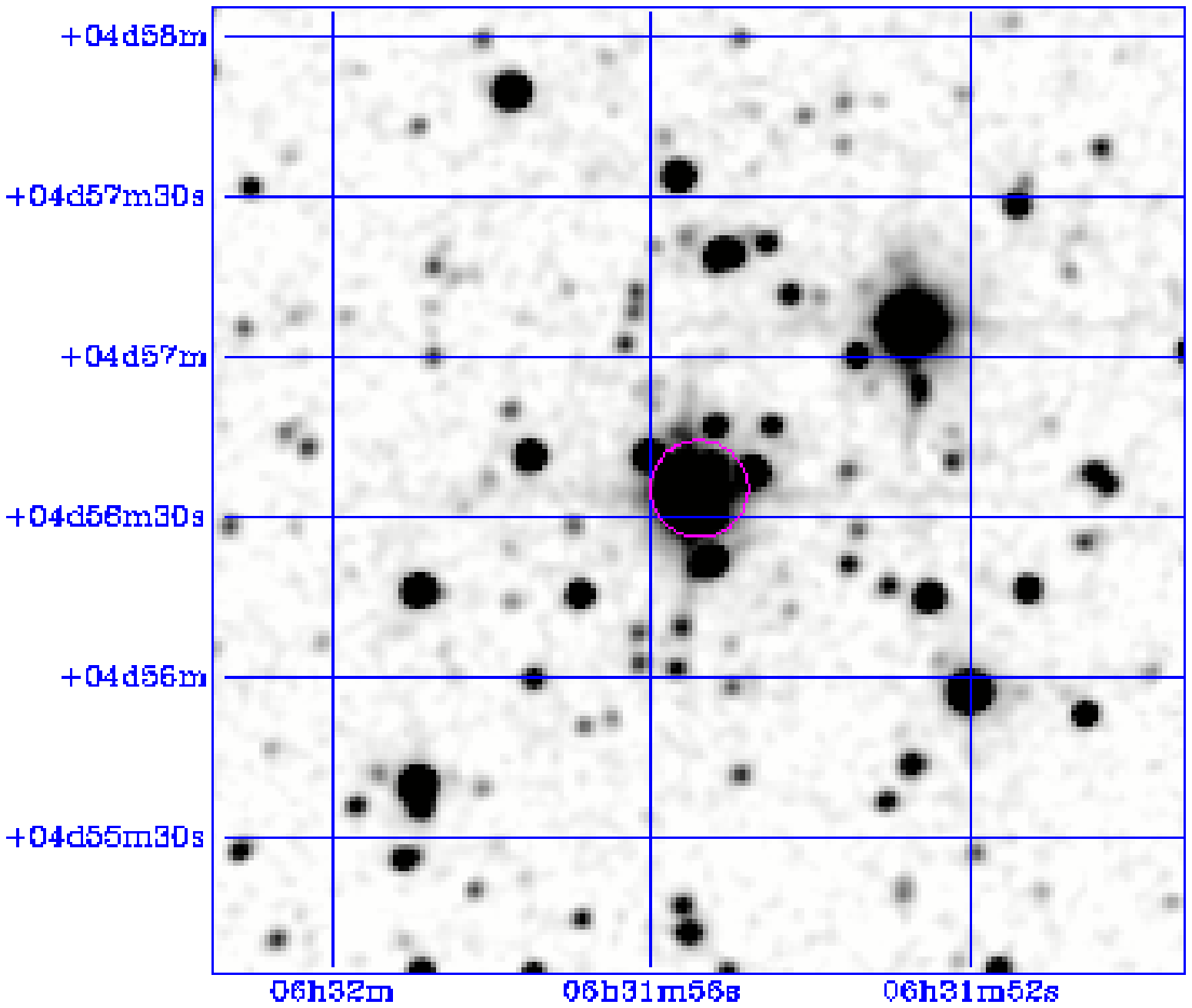}
\end{minipage}\hfill
\caption[]{Top-left panel: $1\degr\times1\degr$ DSS B image of NGC\,2244 and the Rosette 
Nebula, centred on the coordinates of Table~\ref{tab1}. Right: $30\arcmin\times30\arcmin$ 
I image of the same region; circles indicate the cluster radii (Sect.~\ref{struc}) of 
NGC\,2244 (East) and NGC\,2239 (West). Several centres adopted for these clusters are 
identified in the right panel. The K0V star 12\,Mon is indicated by position 'B'. Other 
positions as indicated by the keys in Table~\ref{tab1}. Bottom: \ks\ image of NGC\,2244 
taken from the 2MASS Image Service focusing on the compact core within $3\arcmin\times3\arcmin$. 
Same centre as in the B image. Orientation: North to the top and East to the left. }
\label{fig1}
\end{figure*}

In this paper we apply our set of analytical tools to the 2MASS photometry of the stars 
in the area of NGC\,2244 to derive its fundamental parameters, structure and fraction of 
MS and PMS stars. We also investigate the neighbouring cluster NGC\,2239.

\begin{table*}
\caption[]{Previously adopted centres of NGC\,2244 and NGC\,2239}
\label{tab1}
\renewcommand{\tabcolsep}{2.85mm}
\renewcommand{\arraystretch}{1.25}
\begin{tabular}{cccccccccl}
\hline\hline
Cluster&$\alpha(2000)$&$\delta(2000)$&&$\ell$&$b$&Key&R&&Reference\\
       & (hms)        &($^\circ\,\arcmin\,\arcsec$)&&(\degr)&(\degr)&&(\arcmin)\\
(1)&(2)&(3)&&(4)&(5)&(6)&(7)&&(8)\\
\hline
NGC\,2237&06:31:58.5&$+$04:54:35.7&&206.34&$-2.07$&A&---&&\citet{Zuniga08}\\
NGC\,2244&06:32:18.0&$+$04:52:00.0&&206.42&$-2.02$&B&---&&\citet{SulTif73}\\
NGC\,2244&06:30:36.1&$+$04:58:50.6&&206.12&$-2.34$&C&---&&\citet{Zuniga08}\\
NGC\,2244&06:31:55.0&$+$04:58:30.0&&206.28&$-2.06$&D&---&&WEBDA\\
NGC\,2244/12\,Mon$^\dagger$&06:32:19.2&$+$04:51:21.6&&206.43&$-2.02$&E&$\sim24$&&SIMBAD\\
NGC\,2244&06:31:55.4&$+$04:56:35.3&&206.30&$-2.07$&&$\sim10$&&This work\\
NGC\,2239&06:30:54.0&$+$04:57:00.0&&206.18&$-2.29$&F&$\sim18$&&\citet{SulTif73}\\
NGC\,2239&06:30:57.3&$+$04:58:09.0&&206.17&$-2.27$& &$\sim5$&&This work\\
\hline
\end{tabular}
\begin{list}{Table Notes.}
\item ($\dagger$): 12\,Mon as the centre of NGC\,2244. Col.~7: cluster radius. 
\end{list}
\end{table*}

This paper is organised as follows. In Sect.~\ref{RecAdd} we recall literature data on 
NGC\,2244. In Sect.~\ref{Photom} we describe the 2MASS photometry and compare it
with the available optical data; we also describe the field-star decontamination
algorithm and  build CMDs. In Sect.~\ref{age} we derive cluster fundamental parameters. 
In Sect.~\ref{struc} we derive structural parameters by means of stellar RDPs. In Sect.~\ref{MF} 
we provide estimates of cluster mass. In Sect.~\ref{Discus} we compare the structural 
parameters and dynamical state of the present clusters with those of a sample of nearby 
OCs. Concluding remarks are given in Sect.~\ref{Conclu}.

\section{Previous work on NGC\,2244}
\label{RecAdd}

Several studies on NGC\,2244, especially photometric and spectroscopic ones, are available in 
the literature. 

The WEBDA\footnote{\em obswww.univie.ac.at/webda} database locates the cluster centre at 
$\alpha(2000)=06^h31^m55^s$ and
$\delta(2000)=+04\degr56\arcmin30\arcsec$, and provides a distance from the Sun
$\ds=1.45$\,kpc, reddening $\ebv=0.46$, and an age of 7.9\,Myr.

With UBV photometry, \citet{OgIsh81} obtained $\ebv=0.47$, a 
total to selective extinction ratio $R_V=3.2$, $\ds=1.42$\,kpc, 
4\,Myr of age, and a total mass of $5000\,\ms$. With similar data, \citet{HPV00} derived 
an age of 2\,Myr, $R_V=3.2\pm0.07$, $\ebv=0.44$, and $\ds=1.4\pm0.1$\,kpc.

In a comprehensive study of the Northern Monoceros region, \citet{Perez91} found
an apparent diameter of 24\arcmin, an age within $1.45 - 3.63$\,Myr, $\ds=1.67$\,kpc, 
$\ebv=0.48$, and a total mass of $770\,\ms$ for NGC\,2244.
 
\citet{ParkSung02} found an average $\ebv=0.47\pm0.04$, $R_V=3.1\pm0.2$, and 
$\ds=1.66$\,kpc. By comparing their photometric results with theoretical evolution models, 
they derived a main-sequence turnoff (MSTO) age of 1.9\,Myr and a PMS age spread of about 
6\,Myr. The IMF slope calculated for the mass range $3.2\la m(\ms)\la100$ is flat 
($\chi=-0.3\pm0.1$). 
 
\citet{BergChr02} used optical photometry of X-ray selected stars to estimate an isochrone 
age of 3\,Myr, but significantly younger stars are detected.
 
\citet{Li05} provided updates on the nature of this young OC, including its central position, 
physical scale, and stellar population. They found substructures in NGC\,2244 with 2MASS, in 
particular a companion 6.6\,pc west (in fact NGC\,2239) of the NGC\,2244 centre, and probably 
a major stellar aggregate resembling an arc in structure right below the core. Also, a disc 
fraction of $21\pm3\%$ was estimated for members with masses above $0.8\,\ms$.

In a Chandra study of NGC\,2244, \citet{Wang08} detected over 900 X-ray sources, 77\% of
which having optical or FLAMINGOS NIR counterparts. Their X-ray-selected population is 
estimated to be nearly complete between 0.5 and 3\,\ms. The K-band LFs indicate 
a normal \citet{Salpeter55} IMF for NGC\,2244, which differs from the top-heavy 
one reported in earlier optical studies that lacked a good census of $\la4\,\ms$ stars. 
The X-ray LF indicates a population of $\sim2000$ stars with a spatial distribution 
strongly concentrated around the central O5 star, HD\,46150. The other early O star, 
HD\,46223, has few companions. The cluster's stellar RDP shows two structures: a power-law 
cusp around HD\,46150 extending for $\sim0.7$\,pc, surrounded by an isothermal sphere 
reaching out to 4\,pc with core radius $\rc=1.2$\,pc. This double structure, combined with 
the absence of mass segregation, indicates that NGC\,2244 is not in dynamical equilibrium.  
The fraction of X-ray-selected members with K-band excesses caused by inner protoplanetary 
discs is 6\%, slightly lower than the 10\% disc fraction estimated from FLAMINGOS. The Rosette 
X-ray spectra of OB stars are soft and consistent with the standard model of small-scale shocks 
in the inner wind of a single massive star. 

Recently, \citet{Zuniga08} reviewed the Rosette Complex, in particular they refer to
the position of two OCs, NGC\,2244 at position 'C' in the top-right panel of Fig.~\ref{fig1},
and the other at postion 'A', designated as NGC\,2237. We call attention that their NGC\,2244
actually is in the region of the original NGC\,2239, while NGC\,2237 refers to NGC\,2244,
near the position of HD\,46150. 

\citet{Zuniga08} build a scenario where an expanding H\,II region generated by a large OB 
association interacts with a giant molecular cloud, which harbours a number of embedded 
and open clusters.

The wealth of papers on NGC\,2244 reflects the complex - and, at the same time beautiful - 
nature of the interplay between bright massive stars, faint pre-MS stars and a thinning dust 
shroud, all embodied in a single and relatively nearby object. Fig.~\ref{fig1} illustrates 
this scenario. In the DSS\footnote{Extracted from the Canadian Astronomy Data Centre (CADC), 
at \em http://cadcwww.dao.nrc.ca/} B image (top-left panel) NGC\,2244 emerges from the thin 
dust of the Rosette central part, which is also surrounded by strong gas emission. Indeed, gas 
emission and dust absorption are nearly absent in the XDSS I image (top-right), and especially 
in the 2MASS \ks\ image (bottom). Different centres adopted for NGC\,2244 in previous (mostly 
optical) studies are indicated in the top-right panel. However, when seen in \ks, the cluster 
is highly concentrated on HD\,46150 (bottom panel), suggesting a compact core. A similar centre 
for NGC\,2244 had already been suggested by, e.g. \citet{PTW87}.

The different centres are summarised in Table~\ref{tab1}, which shows some confusion in
the identification of the actual centre of NGC\,2244. As will be discussed in 
Sect.~\ref{struc}, we take as centre the coordinates that present the maximum
stellar density (Fig.~\ref{fig11}) computed within circles of 0.25\arcmin\ in radius, for 
MS and PMS stars taken isolately (Sect.~\ref{CMF}). The resulting coordinates (Table~\ref{tab1}) 
are similar to those given by WEBDA. The same procedure was applied to find the centre of
NGC\,2239 (Sect.~\ref{struc}).

\section{Near-IR and optical photometries compared}
\label{Photom}

Since the Rosette Nebula reaches about 1\degr, it is interesting to compare 
large-scale properties of the optical data with those in the near-IR. 

\begin{figure}
\resizebox{\hsize}{!}{\includegraphics{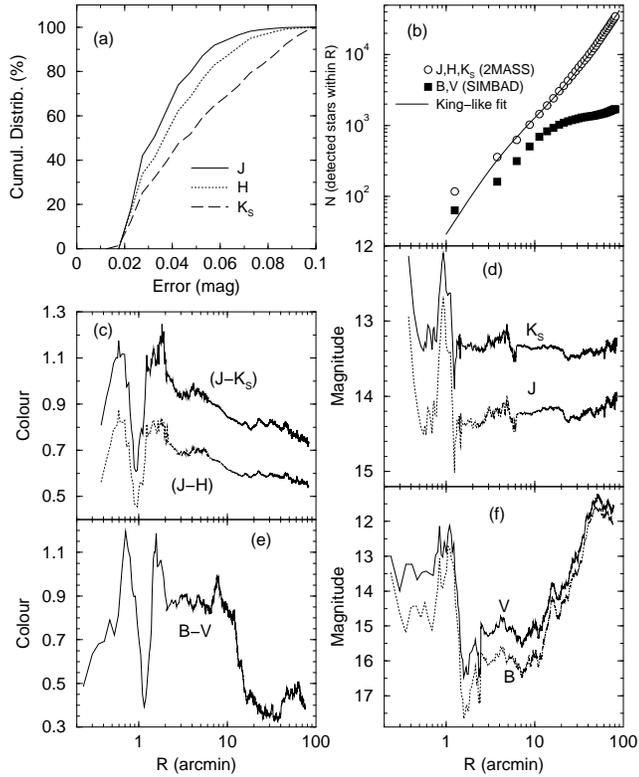}}
\caption[]{Panel (a): cumulative \jj,\ \hh\ and \ks\ photometric error distribution. 
(b): number of stars detected by 2MASS (empty circles) and SIMBAD (filled squares) within 
a given radius; except for the innermost bin, the 2MASS distribution follows a cluster-like
profile (Sect.~\ref{struc}). (c) and (d): spatial dependence of the near-IR colours and
magnitudes. (e) and (f): same for the B and V bands.}
\label{fig2}
\end{figure}

2MASS \jj, \hh, and \ks\ photometry was extracted in a wide circular 
field with VizieR\footnote{\em http://vizier.u-strasbg.fr/viz-bin/VizieR?-source=II/246}. The 
basic condition is that the extraction radius \rx\ should be large enough to allow determination 
of the background level (Sect.~\ref{struc}). We used $\rx=80\arcmin$ (NGC\,2244) and $\rx=30\arcmin$ 
(NGC\,2239), which are considerably larger than the respective cluster radii (Sect.~\ref{struc} and 
Table~\ref{tab3}). In the absence of significant differential absorption (\citealt{BB07}), 
wide extraction areas provide statistics for a consistent colour and magnitude characterisation of 
field stars. For decontamination purposes, comparison fields were extracted within wide rings located 
beyond the cluster radii. As photometric quality constraint, the 2MASS extractions were restricted 
to stars {\em (i)} brighter than the 99.9\% Point Source Catalogue completeness limit\footnote{According 
to the 2MASS Level\,1 Requirement, at {\em http://www.ipac.caltech.edu/2mass/releases/allsky/doc/ }} 
in the cluster direction, and {\em (ii)} with errors in \jj, \hh, and \ks\ lower than 0.1\,mag. 
The 99.9\% completeness limits refer to field stars, and depend on Galactic coordinates. 
Figure~\ref{fig2} (panel a) shows the distribution of uncertainties as a function of magnitude
for the stars in the direction of NGC\,2244. The fraction of stars with \jj, \hh, and \ks\ 
uncertainties lower than 0.05\,mag is $\approx80\%$, $\approx70\%$ and $\approx60\%$, respectively. 
We employ the relations $A_J/A_V=0.276$, $A_H/A_V=0.176$, $A_{K_S}/A_V=0.118$, and $A_J=2.76\times\ejh$ 
(\citealt{DSB2002}), with $R_V=3.1$. They stem from the extinction curve of \citet{Cardelli89}.

The available B and V photometry for NGC\,2244 was taken from 
SIMBAD\footnote{http://simbad.u-starsbg.fr/simbad} within the same extraction radius as that
used for 2MASS. As expected, the number of detected stars at a given radius in the optical 
is significantly lower than in the near-IR (Fig.~\ref{fig2}, panel b). Indeed, the ratio
of the number of stars detected in the near-IR to the optical $N_{NIR}/N_{opt}$ increases
with distance to the cluster centre, being $N_{NIR}/N_{opt}\approx2$ for $R\la10\arcmin$
(approximately the cluster region) and $N_{NIR}/N_{opt}\approx24$ for $R\la80\arcmin$. 
Except for the innermost region, the stellar spatial distribution detected with 2MASS 
follows a cluster-like profile (Sect.~\ref{struc}), while the optical distribution deviates
by a large amount (panel b). One conclusion is that analysis based on star-counts in a
dust-rich region is more realistic in the near-IR than in the optical. Also, panel (b) 
shows that dust is thicker at large radii. This may have introduced biases in some of the
previous optical studies.

The spatial dependence of the colours towards NGC\,2244 is examined in Fig.~\ref{fig2} for 
the 2MASS (panel c) and optical (e) bands. The fiducial lines have been built as running
averages of the raw (observed) data, with 10 points for $R<2\arcmin$, 100 for 
$2\arcmin<R<20\arcmin$, and 1000 for $R>20\arcmin$. Colours in both domains present a similar 
pattern, characterised by a blue core ($R\la2\arcmin$) containing essentially the MS stars.
For larger radii, foreground stars dominate the optical photometry, while the near-IR probes
deeper regions. A similar effect occurs in the average magnitudes (panels d and f).

\begin{figure}
\resizebox{\hsize}{!}{\includegraphics{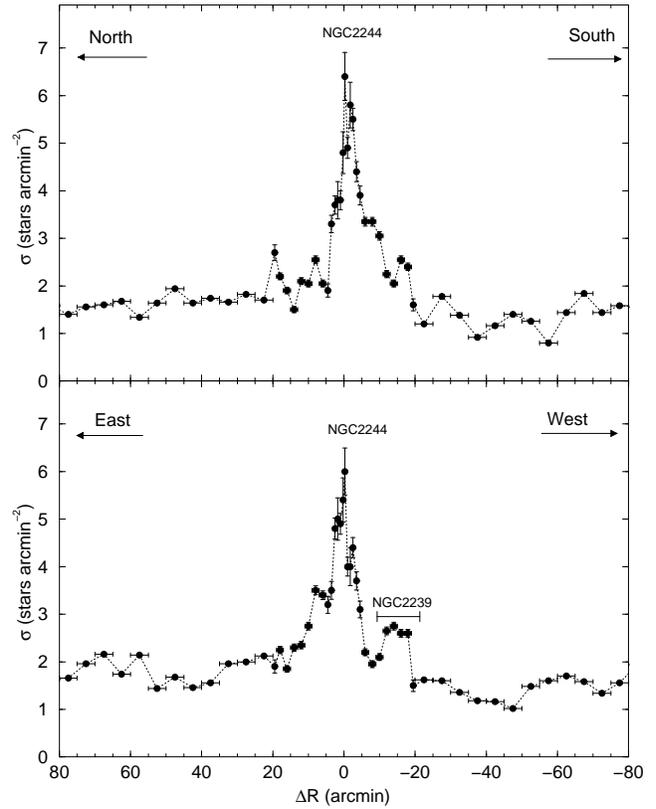}}
\caption[]{Linear extractions centred on NGC\,2244. The extractions are 6\arcmin\ wide in 
both directions. The clusters are $\approx15\arcmin$ apart along the EW direction.}
\label{fig3}
\end{figure}

As suggested by the $\Delta R=6\arcmin$ linear extractions along the N-S and E-W directions
(Fig.~\ref{fig3}), a reasonable 
spatial uniformity level occurs with the 2MASS near-IR photometry. Besides NGC\,2244 itself, 
the next conspicuous bump is caused by NGC\,2239 at $\approx15\arcmin$ to the West. These 
profiles guided the comparison field selection.

\begin{figure}
\resizebox{\hsize}{!}{\includegraphics{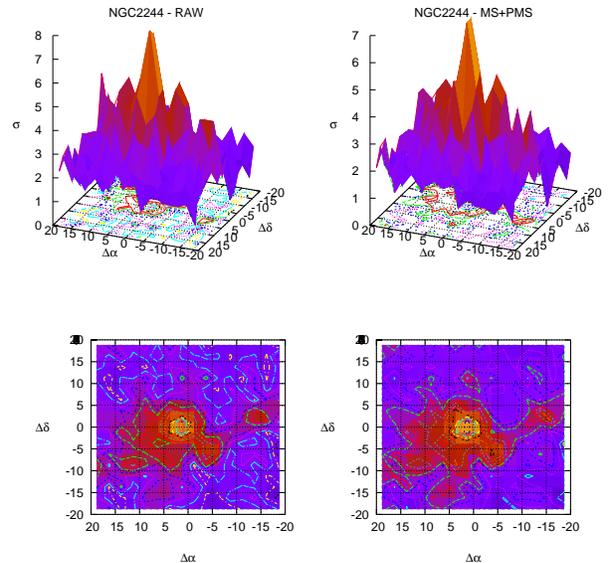}}
\caption[]{Top panels: stellar surface-density $\sigma(\rm stars\ arcmin^{-2})$ of NGC\,2244, 
computed for a mesh size of $2.5\arcmin\times2.5\arcmin$, centred on the coordinates in 
Table~\ref{tab1}. Bottom: the corresponding isopleth surfaces. Left: observed (raw) photometry.
Right: MS and PMS stars selected by means of the colour-magnitude filter (Fig.~\ref{fig6}). 
NGC\,2239 shows up at $\approx15\arcmin$ West of NGC\,2244. $\Delta\alpha$ and 
$\Delta\delta$ in arcmin.}
\label{fig4}
\end{figure}

\begin{figure}
\resizebox{\hsize}{!}{\includegraphics{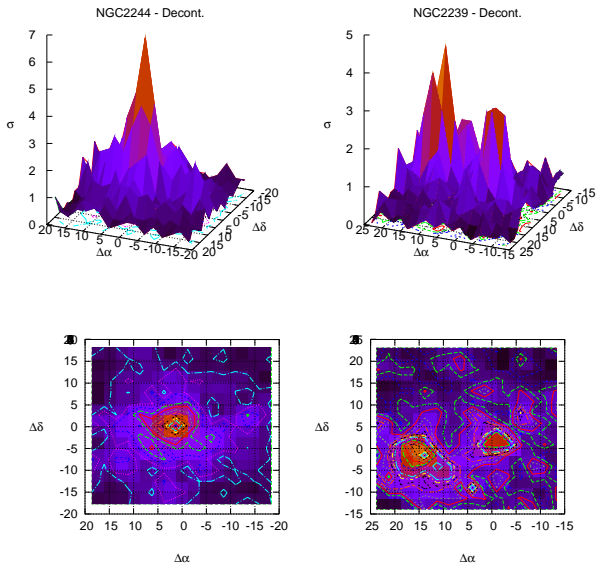}}
\caption[]{Similar to Fig.~\ref{fig4} for the decontaminated photometries of NGC\,2244 
(left panels) and NGC\,2239 (right). NGC\,2239 is the concentration $\approx15\arcmin$ 
to the West of the prominent NGC\,2244.}
\label{fig5}
\end{figure}

Finally, in Fig.~\ref{fig4} (top panels) we show the spatial distribution of the stellar 
surface-density ($\sigma$, in units of $\rm stars\,arcmin^{-2}$) around NGC\,2244, measured 
by 2MASS photometry. The surface density is computed in a rectangular mesh with cells of 
dimensions $2.5\arcmin\times2.5\arcmin$, with meshes reaching total offsets of 
$|\Delta\alpha|=|\Delta\delta|\approx20\arcmin$ with respect to the centre (Table~\ref{tab1}), 
in right ascension and declination. The respective isopleth surfaces are shown in the bottom 
panels, in which NGC\,2239 shows up as a lower concentration at $\approx15\arcmin$ to the West 
of NGC\,2244. Two cases are considered in Fig.~\ref{fig4}, the observed (raw) photometry (left
panels) and the $\rm MS+PMS$ stars taken separately (right) by means of a colour-magnitude filter
(Sect.~\ref{CMF}). Since an important fraction of the contaminant stars are excluded by the
colour-magnitude filter, the surface density distribution of NGC\,2244 (and NGC\,2239) is 
better defined with respect to the surroundings.

\subsection{Colour-magnitude diagrams with 2MASS photometry}
\label{2mass}

\begin{figure}
\resizebox{\hsize}{!}{\includegraphics{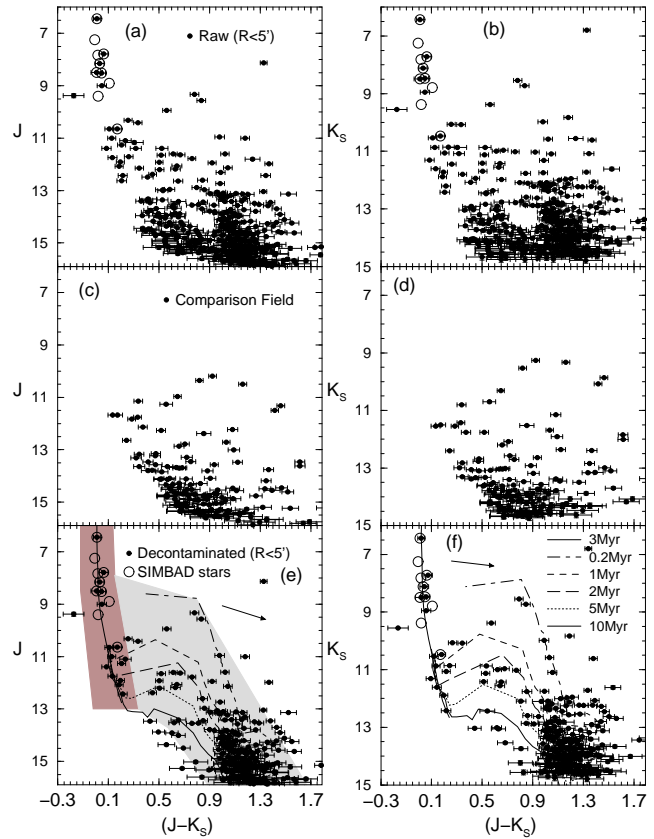}}
\caption[]{2MASS CMDs of NGC\,2244. Top panels: observed photometry extracted from 
the $R<5\arcmin$ region. Middle: equal-area comparison field CMDs, extracted within 
$29.58\arcmin<R<30\arcmin$, including the Mon\,Ob2 association and disc contamination. Bottom panels: 
decontaminated CMDs with the MS fitted by the 3\,Myr Solar-metallicity Padova isochrone. 
PMS tracks of different ages are shown. The shaded polygons correspond to the MS (dark-gray) 
and PMS (light-gray) colour-magnitude filters (Sect.~\ref{CMF}). The bright stars listed in 
SIMBAD are identified as circles; 4 of these are within $5\la R(\arcmin)\la10$. Arrows in the 
bottom panels show the reddening vector computed for $\aV=2$.}
\label{fig6}
\end{figure}

CMDs displaying the $\jj\times\jk$ and $\ks\times\jk$ colours built with the raw photometry 
of NGC\,2244 are shown in Fig.~\ref{fig6} (panels a and b). The sampled region ($R<5\arcmin$)
corresponds to about half the cluster radius (Sect.~\ref{struc}). When qualitatively 
compared with the CMDs extracted from the equal-area comparison field (panels c and d), 
features typical of a very young OC emerge. A relatively vertical and populous MS (at 
$0.0\la\jh,\jk\la0.3$) truncated for stars fainter than $\approx12.5$ (or mass $\la2.8\,\ms$ 
- Sect.~\ref{age}) in both \jj\ and \ks, stand out over the field contamination. 

\begin{figure}
\resizebox{\hsize}{!}{\includegraphics{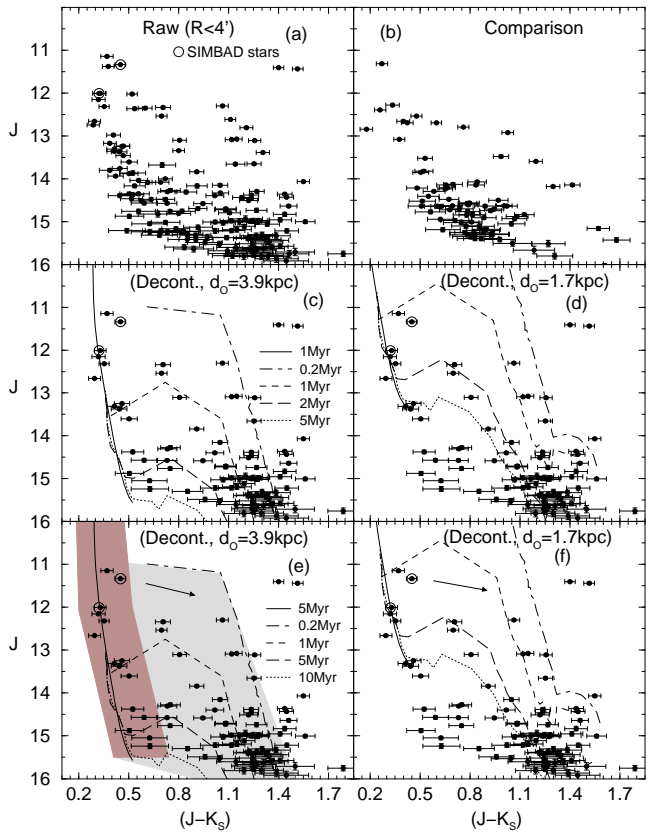}}
\caption[]{Similar to Fig.~\ref{fig6} for the region $R<4\arcmin$ (panel a) of NGC\,2239,
and the equal-area comparison field (b). Panels (c)-(f): decontaminated CMD with different
age/distance from the Sun solutions. Bright stars with SIMBAD optical data are indicated
as open circles. Arrows show the reddening vector computed for $\aV=2$.}
\label{fig7}
\end{figure}

\subsection{Field decontamination}
\label{Decont_CMDs}

As expected of low-Galactic latitude clusters (Table~\ref{tab1}), the stellar surface-density in the 
direction of NGC\,2244 (Fig.~\ref{fig4}) confirms that the field-star contamination, including 
Mon\,OB2 and disc stars, should be taken into account. Further confirmation is provided by the 
qualitative comparison between the CMDs extracted within the cluster and field (Fig.~\ref{fig6}). 
Obviously, the field contribution should be quantified for a better definition of the 
intrinsic CMD morphology.

Although difficult, decontamination is a very important step in the identification 
and characterisation of star clusters. Most of the different approaches 
(e.g. \citealt{Mercer05}) are based essentially on two different premises. The 
first relies on spatial variations of the star-count density, but does not take into account 
CMD evolutionary sequences. Alternatively, stars of an assumed cluster CMD are subtracted 
according to similarity of colour and magnitude with the stars of an equal-area comparison 
field CMD. Together with the present one, these methods are based on photometric properties
only. Ideally, more robust results on membership determination would be obtained if 
another independent parameter, such as the PM of member and comparison field stars, 
is taken into account. However, for PM to be useful the cluster should be 
relatively nearby (e.g. \citealt{AMD03}) and/or to have been observed in widely-apart epochs 
preferentially with high resolution, as for the globular cluster (GC) NGC\,6397
(\citealt{Richer08}). Neither condition is fully satisfied for NGC\,2244, which is relatively 
distant (Sect.~\ref{age}) and was observed by 2MASS in a single epoch. As a consequence, only
about 50\% of the stars within $R=10\arcmin$ of NGC\,2244 have optical PM measured 
(Sect.~\ref{PM}). 

Our decontamination algorithm is fully described in \citet{BB07}, \citet{ProbFSR} and 
\citet{F1603}. For clarity, we provide here only a brief description. The algorithm 
measures the relative number densities of probable field and cluster stars in cubic CMD 
cells with axes along the \jj\ magnitude and the \jh\ and \jk\ colours. It {\em (i)} 
divides the full range of CMD magnitude and colours into a 3D grid, {\em (ii)} estimates 
the number density of field stars in each cell based on the number of comparison field 
stars with similar magnitude and colours as those in the cell, and {\em (iii)} subtracts 
the expected number of field stars from each cell. Input algorithm parameters are the cell 
dimensions $\Delta\jj=1.0$ and $\Delta\jh=\Delta\jk=0.2$; the comparison fields are located 
within $R=30\arcmin-80\arcmin$ (NGC\,2244) and $R=20\arcmin-30\arcmin$ (NGC\,2239). The 
equal-area field extractions (Figs.~\ref{fig6} and \ref{fig7}) should be considered only 
for qualitative comparisons. The decontamination itself uses the large surrounding area 
as described  above. Among other statistical tests, decontaminated CMDs of star clusters 
have integrated $\ns\gg1$ (\citealt{ProbFSR}; \citealt{F1603}), a condition that is fully 
met for the present objects, with $\ns=13.0,~6.5$, respectively for NGC\,2244 and NGC\,2239.

\subsubsection{Decontaminated surface densities and CMDs}
\label{DecOut}

We take the decontaminated surface-density distributions (Fig.~\ref{fig5}) as an efficiency 
indicator. For the present clusters, the central excesses have been significantly 
enhanced with respect to the raw photometry (Fig.~\ref{fig4}), while the residual surface-density 
around the centre has been reduced to a minimum level. By design, the decontamination
depends essentially on the colour-magnitude distribution of stars located in different spatial regions.
The fact that the decontaminated surface-density presents a conspicuous excess only at the assumed 
cluster position implies significant differences among this region and the comparison field, both in 
terms of colour-magnitude and number of stars within the corresponding colour-magnitude bins. This 
meets cluster expectations, which can be characterised by a single-stellar population, projected against 
a Galactic stellar field. 

The decontaminated CMDs are shown in the bottom panels of Figs.~\ref{fig6} and \ref{fig7}. As 
expected, essentially all contamination is removed, leaving stellar sequences typical of mildly
reddened young OCs, with a well-developed MS and a significant population of PMS stars, especially 
in NGC\,2244.

Although in both cases the MS width is rather tight and appears to be dominated by photometric 
errors, we cannot exclude the possibility of differential reddening to account for part of the 
observed spread, especially towards faint stars. To examine this issue we show, in Figs.~\ref{fig6} 
and \ref{fig7} (panels e and f), reddening vectors computed with the 2MASS ratios (Sect.~\ref{2mass}) 
for a visual absorption $\aV=2$, approximately the absorption derived for NGC\,2244 and NGC\,2239 
(Sect.~\ref{age}). Together with the decontaminated CMDs, this experiment shows that differential 
reddening in both clusters is not significant.

We conclude that the qualitative and quantitative expectations of the decontamination algorithm have 
been satisfied by the output. In both cases, the decontaminated photometry presents a relevant excess, 
with respect to the surroundings, in the surface-density distribution (Fig.~\ref{fig5}). 
In addition, field-decontaminated CMDs extracted from the spatial regions where the excesses occur
(Figs.~\ref{fig6} and \ref{fig7}), present statistically significant cluster CMDs.

\subsection{Colour-magnitude filters}
\label{CMF}

To minimise CMD noise, we apply colour-magnitude filters to the raw photometry to exclude stars 
with colours unlike those of the cluster sequence. The filters are wide enough 
to include cluster MS stars and the $1\sigma$ photometric
uncertainties\footnote{Colour-magnitude filter widths should also account for formation or 
dynamical evolution-related effects, such as enhanced fractions of binaries (and other multiple 
systems) towards the central parts of clusters, since such systems tend to widen the MS 
(e.g. \citealt{HT98}; \citealt{Kerber02}; \citealt{BB07}; \citealt{N188}).}. For
very young OCs such as NGC\,2244 and NGC\,2239, we also include filters to account for the
PMS population. The colour-magnitude filters for the present OCs are shown in Figs.~\ref{fig6} 
and \ref{fig7}. 

\subsection{Proper motions}
\label{PM}

Another indication of the star cluster nature of NGC\,2239 is provided by
NOMAD\footnote{{\em http://vizier.u-strasbg.fr/viz-bin/VizieR?-source=I/297.} 
NOMAD is based on the International Celestial Reference System (ICRS) with origin
at the solar system barycenter.} PM data taken for the stars extracted within 
the same spatial region as the 2MASS data.  However, the correspondence between NOMAD and 
2MASS detections is not complete, with $\approx50\%$ of the stars detected with 2MASS 
included in NOMAD, for the colour-magnitude filtered photometry of both NGC\,2244 and 
NGC\,2239. 

\begin{figure}
\resizebox{\hsize}{!}{\includegraphics{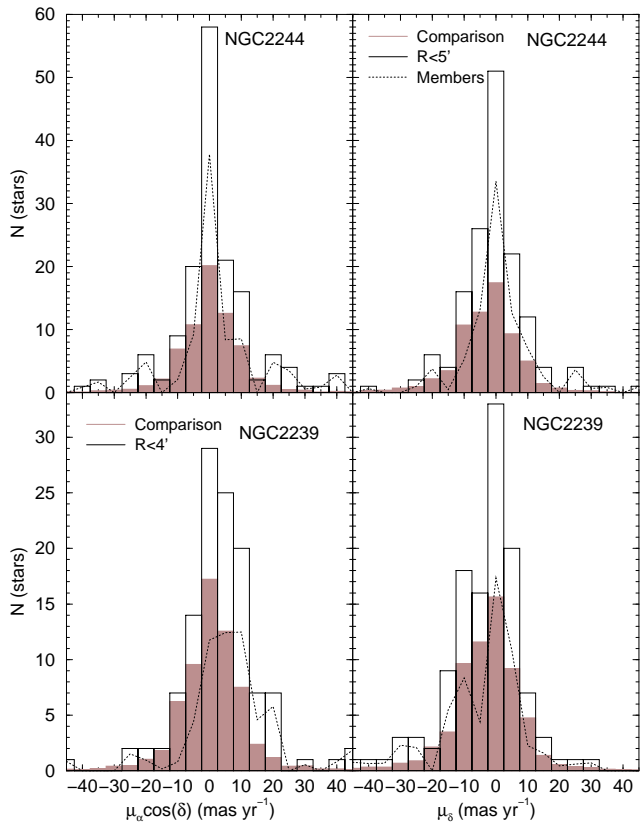}}
\caption{Comparative histograms of the member$+$field (white) and comparison field stars
(shaded), which was scaled to match the projected areas. The intrinsic distributions are
shown by the dashed lines for the extraction $R<5\arcmin$ of NGC\,2244 (top panels) and
$R<4\arcmin$ of NGC\,2239 (bottom).}
\label{fig8}
\end{figure}

In Fig.~\ref{fig8} we show histograms of the right ascension ($\mu_\alpha\cos(\delta)$) and 
declination ($\mu_\delta$) PM components measured for the member$+$field and field stars. We 
use the same cluster and comparison field extractions as those defined for the decontamination 
process (Sect.~\ref{Decont_CMDs}). The field histograms have been normalised to match the 
cluster projected area. Finally, the intrinsic PM distributions are obtained by subtracting 
the normalised field histogram from that of the field$+$members. 

Both NGC\,2244 and NGC\,2239 present conspicuous PM excesses over the field. As expected
of an OC, the intrinsic PM distribution of NGC\,2244 (Fig.~\ref{fig8}, top panels) 
is essentially Gaussian. Although somewhat less defined, a similar conclusion applies to NGC\,2239
(bottom panels). Besides, NGC\,2244 shares essentially the same motion as the disc-field/Mon\,OB2
association, which is consistent 
with a relatively nearby OC. NGC\,2239, on the other hand, appears to be located at a
different distance, especially because of the significant shift in $\mu_\alpha\cos(\delta)$
between member and field stars.

\section{Cluster age, reddening and distance}
\label{age}

The field-decontaminated CMD morphologies (Sect.~\ref{2mass}) can be used to compute 
cluster fundamental parameters. Both NGC\,2244 (Fig.~\ref{fig6}) and NGC\,2239 
(Fig.~\ref{fig7}) present MS and PMS stars that can be used as constraints. We adopt solar
metallicity isochrones because the clusters are young and not far from the Solar circle (see 
below), a region essentially occupied by $[Fe/H]\approx0.0$ OCs (\citealt{Friel95}).

To deal with the MS we use Padova isochrones (\citealt{Girardi2002}) computed with the 2MASS 
\jj, \hh, and \ks\ filters\footnote{{\em http://stev.oapd.inaf.it/cgi-bin/cmd}. These 
isochrones are very similar to the Johnson-Kron-Cousins ones (e.g. \citealt{BesBret88}), with 
differences of at most 0.01 in colour (\citealt{TheoretIsoc}).}. The tracks of \citet{Siess2000} 
are used to characterise the PMS distributions. 

We take $\rs=7.2\pm0.3$\,kpc (\citealt{GCProp}) as the Sun's distance to the Galactic centre
to compute Galactocentric distances, a value derived by means of the GC spatial 
distribution\footnote{Other recent studies gave similar results, e.g. $\rs=7.2\pm0.9$\,kpc 
(\citealt{Eisen03}), $\rs=7.62\pm0.32$\,kpc (\citealt{Eisen05}) and $\rs=7.52\pm0.10$\,kpc
(\citealt{Nishiyama06}), with different approaches.}. 

Historically, different approaches have been used to extract astrophysical parameters from
isochrone fits. The simplest ones are based on a direct comparison of a set
of isochrones with the CMD morphology, while the more sophisticated ones include photometric
uncertainties, binarism, and metallicity variations. Most of these methods are summarised in 
\citet{NJ06}, in which a maximum-likelihood CMD fit method is described. We caution that, 
because of the 2MASS photometric uncertainties for the lower sequences,
a more sophisticated approach for isochrone fitting might lead to an overinterpretation. 

For the above reasons, fits are made {\em by eye}, with the MS and PMS stellar distributions 
as constraint. We also require that, because of the probable presence of binaries, the adopted 
(single-star) MS isochrone should be shifted somewhat to the left of the MS fiducial line, i.e. 
a median line that takes into account the MS spread, including the photometric uncertainties as 
well (e.g. \citealt{N188}, and references therein). In the following we discuss the present clusters 
individually. 

\subsection{NGC\,2244}
\label{NGC2244}

The decontaminated CMD morphology of NGC\,2244 (Fig.~\ref{fig6}) shows a nearly-vertical
MS at $\jj\la13$ and $0.0\la\jk\la0.4$, and a population of low-mass PMS stars at 
$\jj\ga13.5$ and $0.9\la\jk\la1.5$. Taken together, these stellar sequences unambiguously 
characterise a very young OC. 

\begin{figure}
\resizebox{\hsize}{!}{\includegraphics{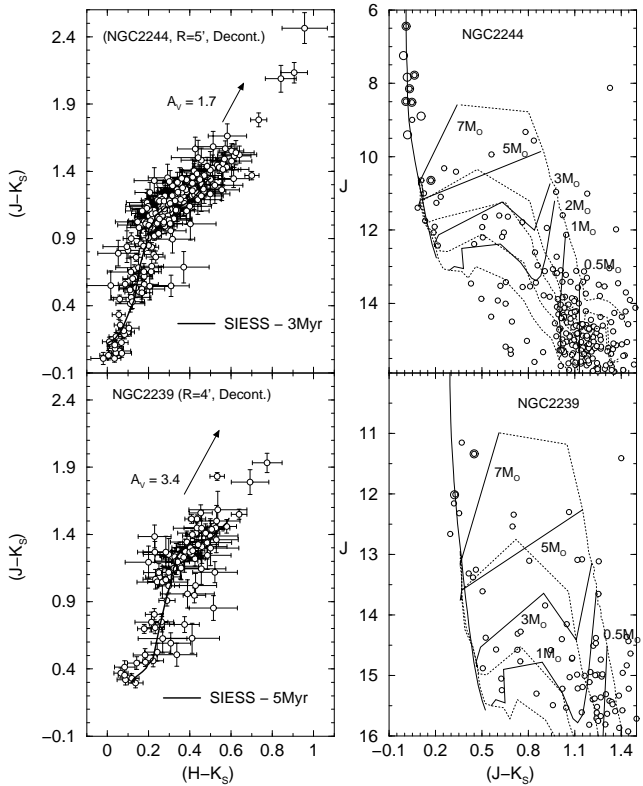}}
\caption[]{Colour-colour diagrams with the decontaminated photometry of NGC\,2244 (top-left
panel) and NGC\,2239 (bottom-left). The 3\,Myr and 5\,Myr \citet{Siess2000} isochrones,set
with the derived reddening values, are used to show the PMS and MS sequences simultaneously.
Reddening vectors for $\aV=1.7,~3.4$ are shown for NGC\,2244 and NGC\,2239, respectively. 
Right panels: evolutionary tracks of PMS stars of different masses are superimposed on the 
decontaminated CMDs of NGC\,2244 (top) and NGC\,2239 (bottom).}
\label{fig9}
\end{figure}

Allowing for photometric uncertainties, acceptable fits to the decontaminated 
MS morphology are obtained with any isochrone with age in the range 1---6\,Myr. The PMS 
stars in Fig.~\ref{fig6} are basically contained within the 0.2\,Myr and 6\,Myr PMS isochrones
(\citealt{Siess2000}), thus implying a similar age range as the MS. Accordingly, we take 
the 3\,Myr isochrone as representative solution. 

With the adopted solution, the fundamental parameters of NGC\,2244 are a near-IR
reddening $\ejh=0.17\pm0.02$ (or $\ebv=0.54\pm0.06$ and $A_V=1.7\pm0.2$), 
an observed and absolute distance moduli $\mMJ=11.5\pm0.2$ and $\mMo=11.03\pm0.21$,
respectively, and a distance from the Sun $\ds=1.6\pm0.2$\,kpc. Thus, for $\rs=7.2$\,kpc, 
the Galactocentric distance of NGC\,2244 is $\dgc=8.7\pm0.2$\,kpc, which puts it $\approx1.5$\,kpc 
outside the Solar circle. This solution is shown in Fig.~\ref{fig6} (panels e and f). The age
spread indicates a non-instantaneous star formation process, similar to what was found for, e.g.
NGC\,4755 (\citealt{N4755}) and NGC\,6611 (\citealt{N6611}).

\subsection{NGC\,2239}
\label{NGC2239}

Although somewhat less-populated than NGC\,2244, the decontaminated CMD of NGC\,2239 also 
presents MS and PMS sequences (Fig.~\ref{fig7}), which suggests a similar age as that of 
NGC\,2244. However, this less-constrained CMD admits alternative age/distance solutions.

Thus, before the CMD fit, we compute the distance of the central bright stars projected
on NGC\,2239. We found 2 bright stars in SIMBAD in common with the 2MASS 
detections. These are the A\,2 star GSC\,00154-01659 ($B=13.32$, $V=12.90$, $\jj=11.336$) and 
the B\,5 star GSC\,00154-02384 ($B=13.42$, $V=13.21$, $\jj=12.014$). With the typical colours 
and magnitudes of stars given in \citet{Binney1998}, the A\,2 star would be located at 
$\ds=1.4$\,kpc, while the B\,5 would be at $\ds=4.3$\,kpc.

Together with young MS and PMS isochrones, both distances are used as starting point to 
search for fundamental parameters. The possible solutions are summarised 
in Fig.~\ref{fig7}, for both the far (left panels) and near (right) distances. We consider 
the MS ages of 1\,Myr (middle panels) and 5\,Myr (bottom). The best fit for these ages was 
for $\ds\approx3.9\pm0.4$\,kpc and $\ds\approx1.7\pm0.3$\,kpc. Within uncertainties, 
they are consistent with those derived for the bright stars. The MS$+$PMS solutions for 
$\ds\approx3.9$\,kpc (Fig.~\ref{fig7}, panels c and f) appear to account for essentially all of the 
decontaminated CMD sequences. Irrespective of age, the near solutions, on the other hand, 
fail to explain several faint stars below the MS.

Considering all available morphological and photometric properties, the best 
overall CMD fit (Fig.~\ref{fig7}, panel e) corresponds to an age $5\pm4$\,Myr, 
$\ejh=0.34\pm0.02$, $\ebv=1.10\pm0.06$, $A_V=3.4\pm0.2$, $\mMJ=13.9\pm0.2$, $\mMo=12.95\pm0.21$,
$\ds=3.9\pm0.4$\,kpc, and $\dgc=10.8\pm0.3$\,kpc, thus $\approx3.6$\,kpc outside the Solar 
circle. Thus, the B\,5 star GSC\,00154-01659 is probably a cluster member, while the A\,2 
star GSC\,00154-01659 is probably in the foreground. We conclude that NGC\,2239 lies 
$\approx2.3$\,kpc in the background of NGC\,2244.

\subsection{Colour-colour diagrams}
\label{2CD}

When transposed to the near-IR colour-colour diagrams $\jk\times\hk$, the age and reddening 
solutions of NGC\,2244 and NGC\,2239 derived above consistently match the field-star 
decontaminated photometry of these OCs (Fig.~\ref{fig9}, left panels). Since they
include PMS stars, we use tracks of \citet{Siess2000} to characterise the 
age of NGC\,2244 ($\sim3$\,Myr) and NGC\,2239 ($\sim5$\,Myr). MS stars lie 
on the blue side of the diagram. Interestingly, only a low fraction of the stars
of these OCs appears to be very reddened. In both cases, the distribution 
suggests a low fraction of disc stars, consistent with the estimates in the range 6-10\%
(\citealt{Wang08}).

\subsection{The fraction of PMS and MS stars}
\label{fracMS_PMS}

Right after formation, most of a cluster's mass is stored in the PMS stars that, eventually, 
shed the dust layers and emerge into the MS (Sect.~\ref{Intro}). Thus, the number of MS and 
PMS stars evolve in opposite directions with cluster age, till all stars are in the MS
after about 30\,Myr (\citealt{N4755}, and references therein). Indeed, NGC\,2244 and NGC\,2239 
present different fractions of MS and PMS stars (Figs.~\ref{fig6} and \ref{fig7}), which is 
consistent with the different ages. 

\begin{figure}
\resizebox{\hsize}{!}{\includegraphics{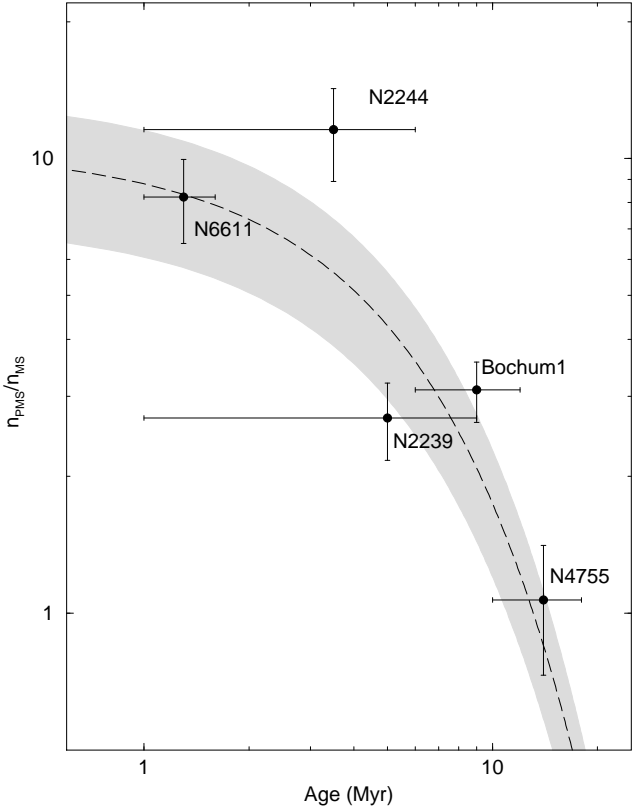}}
\caption[]{The ratio of the number of PMS to MS stars follows an exponential-decay 
function (dashed line) with cluster age: $n_{PMS}/n_{MS}\propto\exp(-age/\tau)$, with
$\tau=5.6\pm1.6$\,Myr. Fit uncertainties ($1\sigma$) are within the shaded region.}
\label{fig10}
\end{figure}

To further explore this issue we compute the ratio of the number of PMS to MS stars 
$f_{PMS/MS}=n_{PMS}/n_{MS}$. Then we examine the age dependence of $f_{PMS/MS}$ for 
the very young OCs, located at a similar distance as NGC\,2244, studied by our group
with the same methods as those employed in NGC\,2244 and NGC\,2239. These conditions
are satisfied by NGC\,6611\footnote{For consistency with the remaining clusters, we counted 
the number of PMS stars brighter than $\jj=16$ in NGC\,6611, since in \citet{N6611} we 
restricted them to $\jj\le15$.} ($1.3\pm0.3$\,Myr, $\ds\approx1.8$\,kpc --- \citealt{N6611}), 
Bochum\,1 ($9\pm3$\,Myr, $\ds\approx2.0$\,kpc --- \citealt{BBD2008}), and NGC\,4755 ($14\pm4$\,Myr, 
$\ds\approx1.8$\,kpc --- \citealt{N4755}). 

Despite the considerable uncertainties in age and the relatively small number of clusters, 
the $f_{PMS/MS}$ ratios shown in Fig.~\ref{fig10} appear to be consistent with the expected 
trend with cluster age. Indeed, we found that the ratios follow the exponential-decay function 
$f_{PMS/MS}\propto e^{-(age/\tau)}$, with the time-scale $\tau=5.6\pm1.6$\,Myr. Thus, it would 
take less than about 30\,Myr for a cluster to retain only $\approx5\%$ (a typical Poisson 
fluctuation) of the original PMS stars. It would be interesting to check this trend - and the 
above time-scale - for a statistically significant sample of young clusters, so that other 
parameters, such as cluster mass, and environmental effects can be considered as well. 

\section{Cluster structure}
\label{struc}

We use the projected stellar RDPs, defined as the stellar number density around the cluster 
centre, to derive structural parameters. To minimise noise, we work with colour-magnitude 
filtered photometry to isolate the MS and PMS stars, which enhances the RDP contrast relative 
to the background, especially in crowded fields (e.g. \citealt{BB07}). However, field stars 
with colours similar to those of the cluster are expected to remain inside the colour-magnitude 
filter, affecting the intrinsic RDP in a way that depends on the relative densities 
of field and cluster stars. The contribution of the residual contamination to the observed 
RDP is statistically evaluated by means of its extension into the field. 

Rings of increasing width with distance from the cluster centre are built to avoid oversampling 
near the centre and undersampling at large radii. The set of ring widths used is
$\Delta\,R=0.25,\ 0.5,\ 1.0,\ 2.0,\ {\rm and}\ 5\arcmin$, respectively for 
$0\arcmin\le R<0.5\arcmin$, $0.5\arcmin\le R<2\arcmin$, $2\arcmin\le R<5\arcmin$, 
$5\arcmin\le R<20\arcmin$, and $R\ge20\arcmin$. The residual background level of each RDP 
corresponds to the average number-density of filtered field stars. 
The $R$ coordinate (and uncertainty) of each ring corresponds to the average position and 
standard deviation of the stars inside the ring.

\begin{figure}
\resizebox{\hsize}{!}{\includegraphics{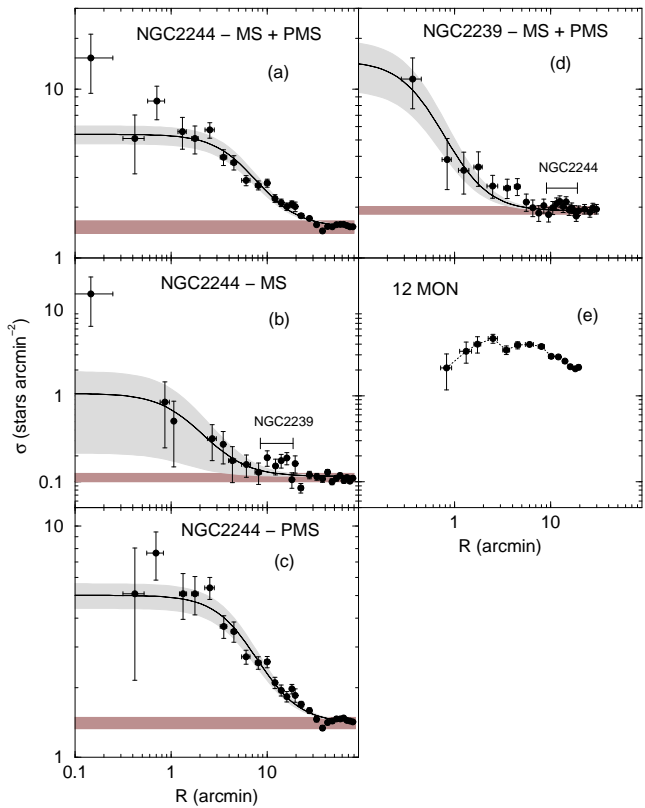}}
\caption[]{Stellar RDPs built with colour-magnitude filtered photometry. Solid line: 
best-fit King-like profile. Horizontal shaded polygon: background. Shaded 
regions: $1\sigma$ King fit uncertainty. Note the central density excess in the RDP of
NGC\,2244 in panels (a) and (b). The RDP with 12\,Mon as centre in shown in (d).}
\label{fig11}
\end{figure}

The colour-magnitude filtered RDPs of the clusters are shown in Fig.~\ref{fig11}. 
As expected, minimisation of the number of non-cluster stars by the colour-magnitude filter 
resulted in RDPs with a high contrast relative to the background. For NGC\,2244 we also 
show the RDPs built with the MS and PMS stars separately (left panels). Interestingly, while
the MS RDP (panel b) has a conspicuous density excess for $R\approx0.15\arcmin$, the PMS 
stars (panel c) are found only for $R\ga0.4\arcmin$. The presence of NGC\,2239 causes a 
bump in the MS RDP of NGC\,2244 (panel b). Similarly, NGC\,2244 shows up in the  
RDP of NGC\,2239 (d). Fig.~\ref{fig11} also shows the RDP 
produced with the star 12\,Mon as centre (panel d). It is clear that
12\,Mon cannot be the centre of NGC\,2244.

\begin{table*}
\caption[]{Derived cluster structural parameters}
\label{tab3}
\renewcommand{\tabcolsep}{2.0mm}
\renewcommand{\arraystretch}{1.25}
\begin{tabular}{cccccccccccc}
\hline\hline
Cluster&$\sigma_{bg}$&$\sigma_0$&\rc&\rl&$\delta_c$&$1\arcmin$&$\sigma_{bg}$&$\sigma_0$&\rc&\rl\\
       &$\rm(*\,\arcmin^{-2})$&$\rm(*\,\arcmin^{-2})$&(\arcmin)&(\arcmin)& &(pc)&
$\rm(*\,pc^{-2})$&$\rm(*\,pc^{-2})$&(pc)&(pc)\\
(1)&(2)&(3)&(4)&(5)&(6)&(7)&(8)&(9)&(10)&(11)\\
\hline
NGC\,2244$^\dagger$&$1.53\pm0.02$&$3.87\pm0.65$&$5.6\pm0.8$&$10.0\pm2.0$&$3.5\pm0.6$&0.466&$7.0\pm0.4$&$17.8\pm3.0$&
   $2.6\pm0.4$&$4.7\pm0.9$\\
   
NGC\,2244$^\ddagger$&$0.12\pm0.01$&$0.95\pm0.43$&$1.2\pm0.9$&$8.0\pm1.0$&$9.2\pm4.1$&0.466&$0.5\pm0.1$&$4.4\pm1.9$&
   $0.6\pm0.4$&$3.7\pm0.5$\\
   
NGC\,2244$^*$&$1.41\pm0.09$&$3.61\pm0.61$&$5.7\pm0.8$&$12.0\pm2.0$&$3.6\pm0.6$&0.466&$6.5\pm0.4$&$16.6\pm2.8$&
   $2.7\pm0.4$&$5.6\pm0.9$\\
   
NGC\,2239$^\dagger$&$1.90\pm0.05$&$12.77\pm4.77$&$0.5\pm0.1$&$5.0\pm1.0$&$7.7\pm2.9$&1.127&$1.5\pm0.1$&$10.0\pm3.7$&
   $0.5\pm0.1$&$5.6\pm1.1$\\
\hline
\end{tabular}
\begin{list}{Table Notes.}
\item RDPs built considering separately the MS$+$PMS ($\dagger$), MS ($\ddagger$) and PMS 
($*$). Core (\rc) and cluster (\rl) radii are given in angular and absolute units. 
Col.~6: cluster/background density contrast parameter ($\delta_c=1+\sigma_0/\sigma_{bg}$), 
measured in the colour-magnitude filtered RDPs. Col.~7: arcmin to parsec scale. 
\end{list}
\end{table*}

Most star clusters have RDPs that follow a well-defined analytical profile e.g., the empirical, 
single mass, modified isothermal spheres of \citet{King66} and \citet{Wilson75}\footnote{It 
assumes a pre-defined stellar distribution function and produces more extended envelopes 
than \citet{King66}.}, and the power law with a core of \citet{EFF87}. Each function is
characterised by a different set of parameters that are related to cluster structure.
For simplicity and considering the error bars of the RDPs in Fig.~\ref{fig11}, we adopt 
the function $\sigma(R)=\sigma_{bg}+\sigma_0/(1+(R/R_c)^2)$, where $\sigma_{bg}$ is the 
residual background density, $\sigma_0$ is the central density of stars, and \rc\ is the 
core radius. It is similar to the function introduced by \cite{King1962} to describe the 
surface brightness profiles in the central parts of GCs. 
To minimise degrees of freedom, $\sigma_0$ and \rc\ are obtained from the fit, while 
$\sigma_{bg}$ is measured in the field. The RDP bins corresponding to the neighbouring 
clusters were ignored in the fit. The best-fit solutions are shown in Fig.~\ref{fig11}, 
and the parameters are given in Table~\ref{tab3}. For absolute comparison with other clusters,
Table~\ref{tab3} gives parameters in absolute units. 

Within uncertainties, the adopted King-like function describes well the colour-magnitude 
filtered RDP of NGC\,2239 (panel d) along the full radius range. The same applies only
for $R\ga1\arcmin$ for the RDP of NGC\,2244. The innermost bin in the MS (and to a lesser 
degree to the MS$+$PMS) RDP (panel b) presents a several $\sigma$ excess over the fit.
This RDP cusp basically corresponds to the detached grouping of stars (with a diameter 
of $\approx0.25\arcmin$) around HD\,46150, seen in Fig.~\ref{fig1} (bottom-right panel).
Our inner RDP shape agrees with that derived by \citet{Wang08} with FLAMINGOS.
In old star clusters, such a central RDP excess can be attributed to a post-core collapse, 
like those detected in some GCs (e.g. \citealt{TKD95}). It has been detected as 
well in Gyr-class OCs, such as, e.g. NGC\,3960 (\citealt{N3960}) and LK\,10 (\citealt{LKstuff}). 
Another very young cluster harbouring such a detached core producing an RDP central cusp 
is NGC\,6823 (\citealt{BBD2008}). Clusters are not expected to dynamically evolve into a 
post-core collapse on short time-scales, and the cusp must have been caused by star-forming 
effects. The compact core within the eroded profile of Bochum\,1 (\citealt{BBD2008}) can be 
a long-lived structure  in young clusters. Consequently, this central cusp in such a young 
cluster as NGC\,2244 suggests a significant deviation from dynamical equilibrium 
(Sect.~\ref{Discus}).

We also estimate the cluster radius (\rl) by visually comparing the cluster RDP and 
background levels, i.e. \rl\ is the distance from the cluster centre where both are 
statistically indistinguishable (e.g. \citealt{DetAnalOCs}, and references therein). 
Most of the cluster stars are contained within $\rl$, which should not be mistaken for 
the tidal radius\footnote{Tidal radii are derived from, e.g. the 3-parameter King-profile 
fit to RDPs (\citealt{StrucPar}), which requires large surrounding fields and adequate 
errors. For instance, in populous and relatively high Galactic latitude OCs such as M\,67, 
NGC\,188, and NGC\,2477, the tidal radii are a factor $\sim2$ larger than \rl\ 
(\citealt{DetAnalOCs}).}. The cluster radii of the present objects are given in 
angular and absolute scales (Table~\ref{tab3}). 

The density contrast parameter $\delta_c=1+\sigma_0/\sigma_{bg}$, which is relatively high
($3.5<\delta_c<9.2$) for the present RDPs, is also given in Table~\ref{tab3}. Since 
$\delta_c$ is measured in colour-magnitude-filtered (lower noise) RDPs, it is usually higher 
than the visual contrast produced by images (e.g. Fig.~\ref{fig1}).


Taken at face value, the core radius of NGC\,2244 (for the MS$+$PMS stars) $\rc\sim2.6$\,pc 
would put it beyond the median value of the distribution derived for a sample of relatively 
nearby OCs by \citet{Piskunov07}. NGC\,2239, on the other hand, falls on the low-\rc\ tail. 
Besides, assuming the relation tidal radius $\sim2\times\rl$ (\citealt{DetAnalOCs}), both 
clusters fall around the median value of the tidal radius distribution.

\section{Cluster mass}
\label{MF}

\begin{figure}
\resizebox{\hsize}{!}{\includegraphics{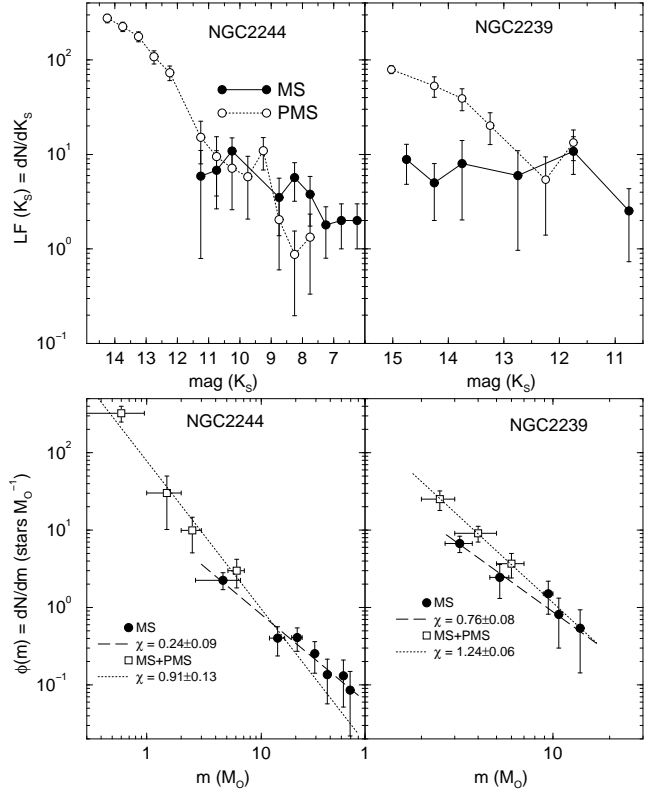}}
\caption[]{Top panels: \ks-luminosity functions of the MS (filled circles) and PMS 
(empty circles) stars. Bottom: mass functions (for the \jj, \hh\ and \ks\ bands combined) of the 
MS (filled circles) and MS$+$PMS (empty squares). Fits of the function 
$\phi(m)\propto m^{-(1+\chi)}$ are shown by the dashed (MS) and dotted (MS$+$PMS) 
lines.}
\label{fig12}
\end{figure}

Both clusters clearly present distinct populations of MS and PMS stars (Figs.~\ref{fig6}
and \ref{fig7}). As the first step to estimate the cluster masses we build the luminosity 
functions (LFs) in the \ks\ band for the MS and PMS stars separately, by means of the 
respective colour-magnitude filters (Sect.~\ref{CMF}). We show them in Fig.~\ref{fig12}, 
where the similar age, different distances and number of members are 
reflected, especially on the different MS and PMS cutoffs. In both cases the PMS LFs
present the expected steep increase towards faint magnitudes (low-mass stars), which
confirms that PMS stars are an important fraction of the members.

For a more objective investigation on the stellar mass distribution we build
the MFs $\left(\phi(m)=\frac{dN}{dm}\right)$ for the current MS stars that, in turn, 
can be used to compute the mass stored in stars. Similarly to the 
RDPs (Sect.~\ref{struc}), we work with colour-magnitude filtered photometry to minimise 
noise. First we build the LF independently for each 2MASS band, both for the cluster 
region ($R<\rl$) and comparison field. The intrinsic LFs are obtained by subtracting the
respective (equal-area) comparison field LF from that of the cluster. The 
intrinsic LFs are transformed into MFs with the mass-luminosity relations obtained 
from the corresponding age and distance from the Sun solutions (Sect.~\ref{age}). The 
final MF is produced by combining the \jj, \hh\ and \ks\ MFs into a single MF. Further 
details on MF construction are given in \citet{FaintOCs}. The effective MS stellar mass 
ranges are $(4.6\pm2.2)\leq m(\ms)\la60$ (NGC\,2244) and $(3.2\pm0.5)\leq m(\ms)\la14$ 
(NGC\,2239). As Fig.~\ref{fig12} (bottom panels) shows, the MS MFs are rather smooth
and present different lower and upper masses, which reflects the lower distance and younger 
age of NGC\,2244 with respect to NGC\,2239.

Since PMS stars are abundant in both clusters, it is important to build their MF as well.
In Fig.~\ref{fig9} (right panels) we show the evolutionary tracks (\citealt{Siess2000}) 
of PMS stars of different masses superimposed on the decontaminated CMDs of NGC\,2244 
and NGC\,2239. It is clear, especially for NGC\,2244, that PMS stars less massive than 
1\,\ms\ are the most abundant component. Similarly to the MS, the PMS MFs are built
with the number of PMS stars among any two tracks in the cluster region 
and comparison field. Finally, we add the MS and PMS MFs to produce the total MF of
each cluster (Fig.~\ref{fig12}).

The number of MS ($n_{MS}$) and PMS ($n_{PMS}$) members in NGC\,2244 (for $R\le\rl$) 
are derived by counting the stars in the background-subtracted colour-magnitude filtered 
photometry. We apply the same approach as above to compute the PMS mass. There are 
$n_{MS}=26\pm3$ and $n_{PMS}=301\pm60$ stars; the corresponding mass values are 
$m_{MS}=389\pm44\,\ms$ and $m_{PMS}=236\pm46\,\ms$ (computed assuming the average mass 
between any two evolutionary tracks in Fig.~\ref{fig9}). Thus, the total stellar mass of
NGC\,2244 is $m_{MS+PMS}\approx625\,\ms$, which agrees with the 770\,\ms\ mass estimated 
by \citet{Perez91}. We note that this value is about 10\% of the 
mass estimated by \citet{OgIsh81} for NGC\,2244. However, this difference may arise from
the present detailed analysis - especially the decontamination and the separation of MS and 
PMS stars in the construction of the cluster MF. The same analysis applied to NGC\,2239 
yields $n_{MS}=26\pm3$, $n_{PMS}=70\pm11$, $m_{MS}=141\pm16\,\ms$, $m_{PMS}=160\pm32\,\ms$, 
and the total stellar mass $m_{MS+PMS}\approx301\,\ms$, about half the mass of NGC\,2244. 

Considering the MS stars isolately, the MFs can be well represented by the function 
$\phi(m)\propto m^{-(1+\chi)}$, with the slopes $\chi=0.24\pm0.09$ and $\chi=0.76\pm0.08$,
respectively for NGC\,2244 and NGC\,2239. Both values are flatter than the $\chi=1.35$ of 
\citet{Salpeter55} initial mass function (IMF). A flat MF slope was also found for
NGC\,2244 by \citet{ParkSung02}. However, when the MS and PMS stars are taken together, 
the slopes become steeper, $\chi=0.91\pm0.13$ and $\chi=1.24\pm0.06$. While within 
uncertainties the total MF of NGC\,2239 is comparable to the \citet{Salpeter55} IMF, 
the MF of NGC\,2244 remains somewhat flatter, but still consistent with the
conclusions of \citet{Wang08}. 

\section{Discussion} 
\label{Discus}

Constrained by isochrone fits (Sect.~\ref{age}), we could derive fundamental and structural 
parameters of the young OCs NGC\,2244 and NGC\,2239, part of them for the first time. We use 
them to compare some of their properties with those of well-studied OCs.

\subsection{Diagnostic diagrams}
\label{DD}

We further investigate the nature of NGC\,2244 and NGC\,2239 with diagrams that examine 
relations among astrophysical parameters of OCs in different environments. They were 
introduced by \citet{DetAnalOCs}. As reference sample we use some bright nearby 
OCs (\citealt{DetAnalOCs}; \citealt{N4755}), and a group of OCs projected towards the central 
parts of the Galaxy (\citealt{BB07}). Also included are the young OCs NGC\,6611 with the
age $\sim1.3$\,Myr (\citealt{N6611}), NGC\,6823 with $\sim4$\,Myr and Bochum\,1 with
$\sim9$\,Myr (\citealt{BBD2008}). NGC\,6611 and NGC\,6823 serve as comparison with 
gravitationally bound objects of similar age, while Bochum\,1 is a star cluster fossil 
remain that might be dynamically evolving into an OB association. The full sample of 
comparison OCs is characterised by ages in the range $\sim1.3$\,Myr to $\sim7$\,Gyr, and 
Galactocentric distances within $\rm5.8\la\dgc(kpc)\la8.1$. Their parameters have been 
obtained following the same prescriptions as those for NGC\,2244 and NGC\,2239. 

\begin{figure}
\resizebox{\hsize}{!}{\includegraphics{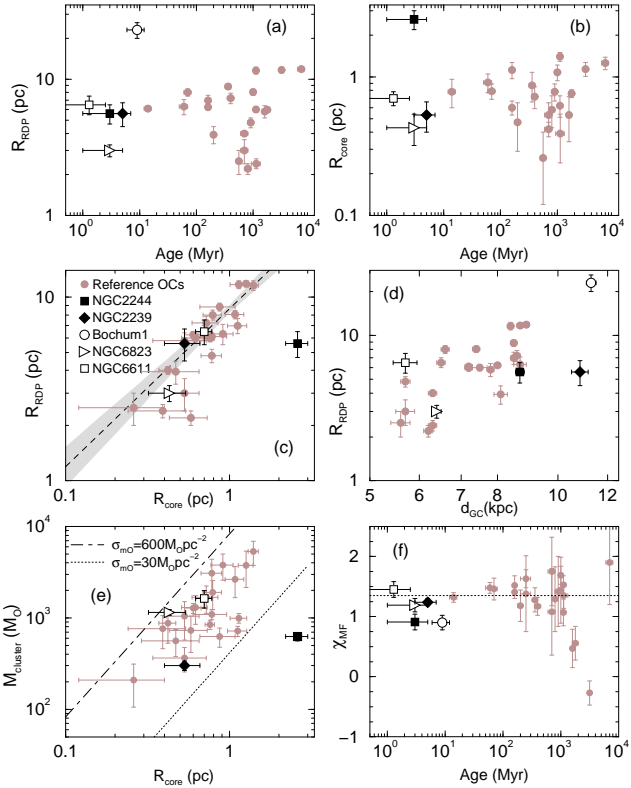}}
\caption{Diagrams dealing with astrophysical parameters of OCs. Gray-shaded circles:
reference OCs. The young star clusters NGC\,6611, NGC\,6823 and Bochum\,1 are indicated
for comparison purposes. Analytical relations in panels (c) and (e) are discussed in the 
text. Dotted line in panel (f) shows \citet{Salpeter55} IMF slope $\chi=1.35$.}
\label{fig13}
\end{figure}

The diagrams are shown in Fig.~\ref{fig13}, where panels (a) and (b) examine the dependence 
of cluster (\rl) and core (\rc) radii on cluster age, respectively. Most of the 
small-radius OCs (especially in \rl) occur at an age $\sim0.5-1$\,Gyr, the typical time-scale of 
OC disruption processes near the Solar circle (e.g. \citealt{Bergond2001}; \citealt{Lamers05}).
Both NGC\,2244 and NGC\,2239 present a cluster radius comparable to that of NGC\,6611 (and in
general equivalent to other young OCs). The same applies to the core radius of NGC\,2239. 
NGC\,2244, on the other hand, has an \rc\ too large when compared to the reference OCs.

Core and cluster radii of the reference OCs follow the relation $\rl=(8.9\pm0.3)\times
R_{\rm C}^{(1.0\pm0.1)}$ (panel c)\footnote{Similar relations were also found by 
\citet{Nilakshi02}, \citet{Sharma06}, and \citet{MacNie07}.}, suggesting a similar scaling for 
both kinds of radii. While NGC\,2239 fits tightly in that relation, NGC\,2244 deviates again 
probably because of the exceeding cluster radius. A dependence of OC size on Galactocentric 
distance is suggested by panel (d), as discussed by \citet{Lynga82} and \citet{Tad2002}. 
While NGC\,2244 follows the trend, NGC\,2239 deviates somewhat. This relation may be partly 
primordial, in the sense that the high molecular gas density in central Galactic regions may 
have produced small clusters (e.g. \citealt{vdB91}). After formation, mass loss due to stellar 
and dynamical evolution (e.g. mass segregation and evaporation), together with tidal interactions 
with the Galactic potential and giant molecular clouds, also contribute to the depletion of star 
clusters, especially the low-mass and centrally located ones.

When the mass-density radial distribution follows a King-like profile (e.g. \citealt{OldOCs};
\citealt{StrucPar}; \citealt{PlaNeb}), the cluster mass inside \rl\ can be computed as 
a function of the core radius (\rc) and the central mass-surface density ($\sigma_{M0}$), 
$\rm M_{clus}=\pi\,R^2_{C}\sigma_{M0}\ln\left[1+\left(\rl/\rc\right)^2\right]$. With the 
above relation (panel c) between \rc\ and \rl, this equation becomes 
$\rm M_{clus}\approx13.8\sigma_{M0}\,R^2_{C}$. The observed relation of core radius and 
cluster mass is examined in panel (e). The reference OCs, together with NGC\,2239 are contained 
within King-like distributions with central mass densities within 
$\rm30\la\sigma_{M0}\,(\ms\,pc^{-2})\la600$. NGC\,2244 appears to have too big a core radius
for its total mass. 

Finally, when the total (MS$+$PMS) MF slope is considered (panel f), NGC\,2239 and especially 
NGC\,2244, appear to have MFs flatter than those of similarly young OCs. On the other hand, 
their slopes are equivalent to those derived for some old OCs in the reference sample. 
In general, flat MFs reflect advanced dynamical evolution (e.g. \citealt{DetAnalOCs}).

What follows from the above analysis is that while NGC\,2239 appears to be characterised 
by parameters of a typical young OC, NGC\,2244, on the other hand, exhibits evidence of 
a system far from dynamical equilibrium, which agrees with \citet{CdGZ07} and \citet{Wang08}.

\section{Summary and conclusions}
\label{Conclu}

In the present paper we employ the wide-field and near-IR depth provided by 2MASS 
to focus on the Rosette Nebula cluster NGC\,2244 and the nearby projected OC NGC\,2239. 
Our approach relies essentially  
on field-star decontaminated 2MASS photometry, which enhances cluster CMD evolutionary 
sequences and stellar radial density profiles, producing more constrained fundamental 
and structural parameters.

Previous studies were mostly based on optical photometry and/or near-IR with small 
angular fields. However, 2MASS can still provide additional insight (Sect.~\ref{Photom}). 
The set of tools developed by our group allowed to unambiguously isolate MS and PMS 
stars that, in turn, resulted in well-defined CMDs, RDPs and mass functions. In addition, 
we explore proper motion properties to investigate the other cluster in the area,
NGC\,2239.

Taken together, the (decontaminated) MS and PMS sequences of NGC\,2244 provided an 
age range 1---6\,Myr, an absorption $\aV=1.7\pm0.2$, and a distance from the Sun 
$\ds=1.6\pm0.2$\,kpc ($\approx1.5$\,kpc outside the Solar circle). These parameters
are consistent with most of the previous estimates. The (MS$+$PMS) MF slope $\chi=0.91\pm0.13$ 
is somewhat flatter than the \citet{Salpeter55} IMF. With a total (MS$+$PMS) stellar 
mass of $m_{MS+PMS}\sim625\,\ms$ derived in the present work, NGC\,2244 is not as 
massive as previously estimated (Sect.~\ref{RecAdd}). The King-like profile fit to 
the (MS$+$PMS) stellar RDP was obtained with a core radius $\rc\approx5.6\arcmin\approx2.6$\,pc;
the corresponding cluster radius is $\rl\approx10\arcmin\approx4.7$\,pc. Compared
to a set of well-known OCs, the core radius of NGC\,2244 appears to be abnormally big,
which puts it at unusual loci in the structural/dynamical diagnostic-diagrams 
(Sect.~\ref{DD}). NGC\,2244 has a central cusp that cannot be fitted by e.g., a King's law.
This cusp is probably due to a star-formation effect, and not the product of dynamical
evolution. We conclude that NGC\,2244 is not in dynamical equilibrium, consistent
with \citet{CdGZ07} and \citet{Wang08}.

NGC\,2239 is a low-mass ($m_{MS+PMS}\approx301\,\ms$), young ($5\pm4$\,Myr)
and somewhat more absorbed ($\aV=3.4\pm0.2$) OC, at $\ds=3.9\pm0.4$\,kpc, 
thus in the background of 
NGC\,2244. Structurally, its RDP can be represented by a King-like profile with 
$\rc\approx0.5\arcmin\approx0.5$\,pc and $\rl\approx5.0\arcmin\approx5.6$\,pc. With 
$\chi=1.24\pm0.06$, its composite MS$+$PMS MF slope is essentially Salpeter's IMF. These
parameters characterise an average young OC, as compared to the reference nearby OCs 
(Sect.~\ref{DD}).

While NGC\,2239 is a {\em normal} young OC with MS and PMS stars distributed according to a 
cluster RDP, NGC\,2244 appears to be another example, like Bochum\,1 (\citealt{BBD2008}), of 
an open cluster doomed to dissolution in a few $10^7$\,yr. The present work shows the
importance of field-star decontamination and wide-field extractions to get the best stellar
statistics and to produce high-quality CMDs and RDPs.

\section*{Acknowledgements}
We thank the referee, Dr. M. P\'erez, for comments.
We acknowledge support from the Brazilian Institution CNPq.
This publication makes use of data products from the Two Micron All Sky Survey, which
is a joint project of the University of Massachusetts and the Infrared Processing and
Analysis Centre/California Institute of Technology, funded by the National Aeronautics
and Space Administration and the National Science Foundation. This research has made 
use of the WEBDA database, operated at the Institute for Astronomy of the University
of Vienna.

\label{lastpage}
\end{document}